\documentclass[5p]{elsarticle}
\usepackage{graphicx}
\usepackage{multirow}%
\usepackage{amsmath,amssymb,amsfonts}%
\usepackage{amsthm,mathrsfs}%
\usepackage[title]{appendix}%
\usepackage{xcolor}%
\usepackage{textcomp}%
\usepackage{manyfoot}%
\usepackage{booktabs}%
\usepackage{algorithm}%
\usepackage{algorithmicx}%
\usepackage{algpseudocode}%
\usepackage{listings}%
\usepackage{colortbl}
\usepackage{url}
\usepackage{hyperref}
\journal{Pattern Recognition Letters}

\begin{document}

\begin{frontmatter}
\title{Robust Persian Digit Recognition in Noisy Environments Using Hybrid CNN-BiGRU Model}

\author{Ali Nasr-Esfahani$^1$} 
\author{Mehdi Bekrani$^{*1}$} 
\author{Roozbeh Rajabi$^2$} 

\address{$^1$Faculty of Electrical and Computer Engineering, Qom University of Technology, Iran}
\address{$^2$Hyperspectral Imaging Laboratory, University of Alaska Fairbanks, Fairbanks, AK, USA}

%

\begin{abstract}
Artificial intelligence (AI) has significantly advanced speech recognition applications. However, many existing neural network-based methods struggle with noise, reducing accuracy in real-world environments.
This study addresses isolated spoken Persian digit recognition (zero to nine) under noisy conditions, particularly for phonetically similar numbers.
A hybrid model combining residual convolutional neural networks and bidirectional gated recurrent units (BiGRU) is proposed, utilizing word units instead of phoneme units for speaker-independent recognition.
The FARSDIGIT1 dataset, augmented with various approaches, is processed using Mel-Frequency Cepstral Coefficients (MFCC) for feature extraction. 
Experimental results demonstrate the model's effectiveness, achieving 98.53\%, 96.10\%, and 95.92\% accuracy on training, validation, and test sets, respectively. In noisy conditions, the proposed approach improves recognition by 26.88\% over phoneme unit-based LSTM models and surpasses the Mel-scale Two Dimension Root Cepstrum Coefficients (MTDRCC) feature extraction technique along with MLP model (MTDRCC+MLP) by 7.61\%.
\end{abstract}


\end{frontmatter}

\section{Introduction}\label{sec1}

Artificial intelligence (AI) has become widely used in signal processing, including audio, speech, image processing, and machine vision~\cite{Dhanjal1,Veisi2020,shahidi2022deep}. One common human communication method is through numbers, which are essential in daily life for tasks like financial transactions, phone numbers, and passwords. AI has been used to recognize spoken numbers, eliminating the need for physical inputs, leading to significant interest in speech and digit recognition.

Research on spoken number recognition has been ongoing. Homayounpour et al.~\cite{homayounpour2003rec} reviewed methods using hidden Markov models (HMMs) and MLP neural networks for Persian number recognition, achieving 99.1\% accuracy for discrete and 83.7\% for continuous numbers in the FARSDIGIT1 database~\cite{data_set}. In 2008, Hierarchical Temporal Memory (HTM) was applied to isolated digit recognition~\cite{van2008spoken}, and a Gaussian mixture model (GMM) classifier with and Delta-Delta Mel-Frequency Cepstral Coefficients (MFCC) for feature
extraction achieved 99.3\% accuracy in Arabic digit recognition~\cite{hammami2013spoken}.

In recent years, deep neural networks (DNNs) have gained significant interest in speech recognition. Danashri et al.~\cite{dhanashri2017isolated} used a DNN with a Deep Belief Network (DBN), achieving 86.06\% accuracy for English numbers from the TIDIGIT database~\cite{leonard1993tidigits}. Zada et al.~\cite{zada2020pashto} utilized a Convolutional Neural Network (CNN) for Pashto number recognition, achieving an accuracy of 84.17\%. The model featured four deep convolutional layers and a maximum pooling layer. However, the effect of noise on recognition accuracy remains largely unexplored. 

In~\cite{tabibian2021},  an LSTM-based network was presented that  categorizes discrete numbers into groups with similar phonetic characteristics and trains an LSTM neural network for each class, independently. It achieved 91.7\% accuracy in noise-free conditions but dropped to an average of  69.22\% when various noises are added to the audio data.

More recently, the combination of multi layer perceptron (MLP) and Mel-scale Two Dimension Root Cepstrum Coefficients (MTDRCC) feature extraction method was used for number recognition in noisy conditions, achieving 98.85\% accuracy in noise-free conditions and 88.49\% in noisy ones~\cite{Z2N}, for Persian numbers. Viriri et al.~\cite{oruh2022long} combined recurrent neural network (RNN) and LSTM, achieving 99\% accuracy on English numbers, while Sotisna et al.~\cite{amadeus2022digit} used transfer learning networks, like AlexNet and GoogleNet, reporting lower recognition rates, 72\% for AlexNet and 66\% for GoogleNet. In 2023, a hybrid deep CNN model for Bengali digit recognition, utilizing a unique hybrid feature extraction, including MFCC, Spectral Sub-band Energy (SSE), and Log Spectral Sub-band Energy (LSSE), achieved 98.52\% accuracy~\cite{paul2023hybrid}.

To address limited data for many non-English languages, data augmentation has proven effective. Lunas et al.\cite{lounnas2022} applied techniques like adding white noise and altering sound length, using a Markov model for recognition. Tom Ko et al.\cite{ko2017study} expanded a 300-hour dataset to 900 hours with noise and room simulation techniques, employing MFCC and BiLSTM for recognition.

This paper presents a novel Persian digit recognition approach, addressing linguistic and recognition challenges. Key aspects include:

\begin{itemize}  
  \setlength{\itemsep}{0pt}  
  \setlength{\parskip}{0pt}  
  \item \textbf{Data Augmentation:} Improves robustness with limited data.  
  \item \textbf{MFCC Features:} Extracts relevant speech characteristics.  
  \item \textbf{Word Units:} Enhances accuracy over phoneme-based methods.  
  \item \textbf{Hybrid Network:} Combines Residual CNN and BiGRU for superior performance.  
\end{itemize}

In addition, our method investigates the effects of various noises, including monotonous horns, nature sounds, vehicle movement sounds, humming sounds, and factory sounds, on the speech recognition performance.

The paper is organized as follows: Section~\ref{sec2} covers the data generation method. Section~\ref{sec3} discusses the DNN architecture. Section~\ref{sec4} presents experimental results, and Section~\ref{sec5} concludes with key takeaways  from the study.

\section{Datasets}\label{sec2}

This study utilizes recordings from 51 speakers in the FARSDIGIT1 database~\cite{data_set}, containing discrete and continuous Persian numbers (zero to nine). The recordings, captured over telephone lines (both intra- and inter-city), have an SNR of approximately 8.8 dB and a sampling rate of 11025 Hz. The dataset includes 31 male speakers (ages 12–61) and 20 female speakers (ages 14–52). Each speaker recorded numbers in one to two sessions, 7 to 30 days apart, with 10 repetitions per number, totaling 510 audio data samples per digit.  

To prevent neural network overfitting, we apply the data augmentation on the dataset using:  
\begin{itemize}
  \setlength{\itemsep}{0pt}  
  \setlength{\parskip}{0pt}    
  \item Sound Speed Variation~\cite{lounnas2022}: Adjusting playback speed.  
  \item Reverb Filter~\cite{ko2017study}: Simulating different acoustic environments.  
  \item Background Noise~\cite{hartmann2016two}: Adding various ambient sounds.  
  \item Hall Environment Simulation~\cite{park2019specaugment}: Introducing additional reverberation effects.  
\end{itemize}

The data augmentation methods were applied with different probabilities: 70\% for noise, 15\% for speed changes, and 7.5\% each for reverb and hall simulation. Noise sources included horns, nature sounds, vehicle engines, buzzing, and industrial sounds, with SNR levels of 0, 5, 10, 15, and 20 dB.  

These probabilities were chosen to ensure diverse audio conditions: 70\% noise to introduce varied SNR levels, 15\% speed changes to explore audio variations, and 7.5\% each for reverb and hall simulation due to their similar shape features. This distribution supports a balanced augmentation strategy for comprehensive dataset enrichment.

\begin{figure}[h]  
\centering  
\includegraphics[width=0.48\textwidth]{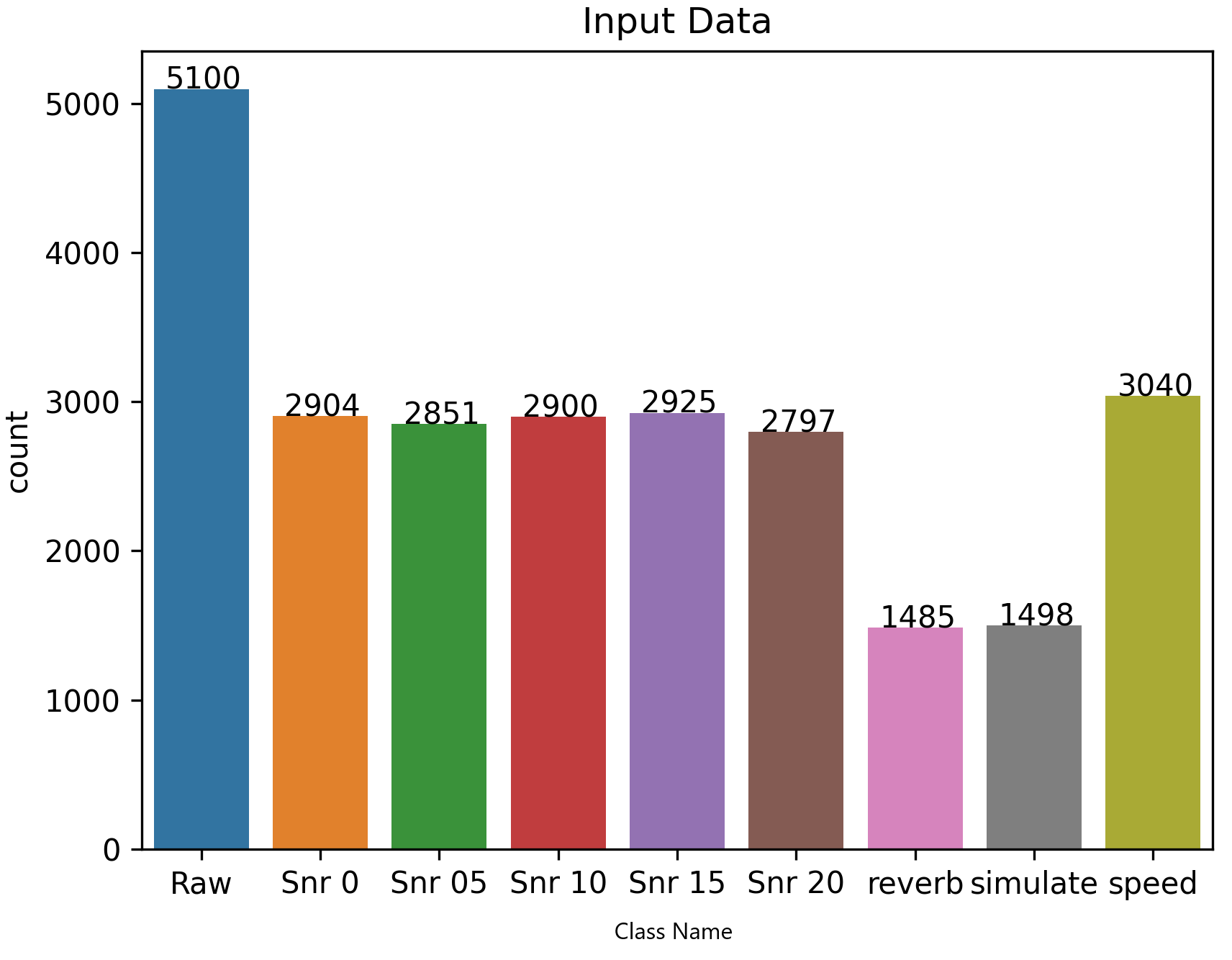}  
\caption{Input data count after data augmentation.}\label{fig1}  
\end{figure}

Applying augmentation increased the dataset size fivefold, resulting in 25,500 data samples (2,550 per digit). Figure~\ref{fig1} illustrates the distribution of augmented data.  

This expanded dataset enhances DNN training, ensuring robustness in diverse real-world noise conditions.  

\section{The Proposed Spoken Digit Recognition Method}\label{sec3}

The proposed method consists of three main stages:
\begin{itemize}
  \setlength{\itemsep}{0pt}  
  \setlength{\parskip}{0pt}  
  \item Data augmentation technique described in Section~\ref{sec2}.
  \item Speech feature extraction using the MFCC technique.
  \item The proposed neural network architecture.
\end{itemize}

\subsection{Speech Feature Extraction Using the MFCC Technique}\label{subsec2}
Feature extraction is crucial in speech recognition, as raw audio contains noise and irrelevant parameters that affect accuracy. Common feature extraction techniques include~\cite{dave2013feature}:

\begin{enumerate}
  \item MFCCs: Capture both spectral and temporal speech characteristics.
  \item Perceptual Linear Prediction: Similar to MFCCs but models human auditory perception with non-linearity.
  \item Mel-Frequency Discrete Wavelet Coefficients: Applies wavelet transform to the speech signal in the mel-frequency domain.
  \item Spectrogram: 2D representation of frequency changes over time.
\end{enumerate}

MFCC stands as the primary choice for speech feature extraction due to its efficiency, noise resilience, and ability to emulate human auditory perception.  MFCCs excel in capturing both spectral and temporal characteristics of speech, while reducing the dimensionality of the data, thus offering an efficient representation of speech signals~\cite{dave2013feature}.

In the proposed method, to account for and mitigate the impact of facet similarities on the recognition rate, the input data is first separated based on words and numbers and the MFCC technique is subsequently used to extract features from segmented speech signals instead of using raw audio signals.
%

\begin{figure}[h]%
\centering
\includegraphics[width=0.45\textwidth]{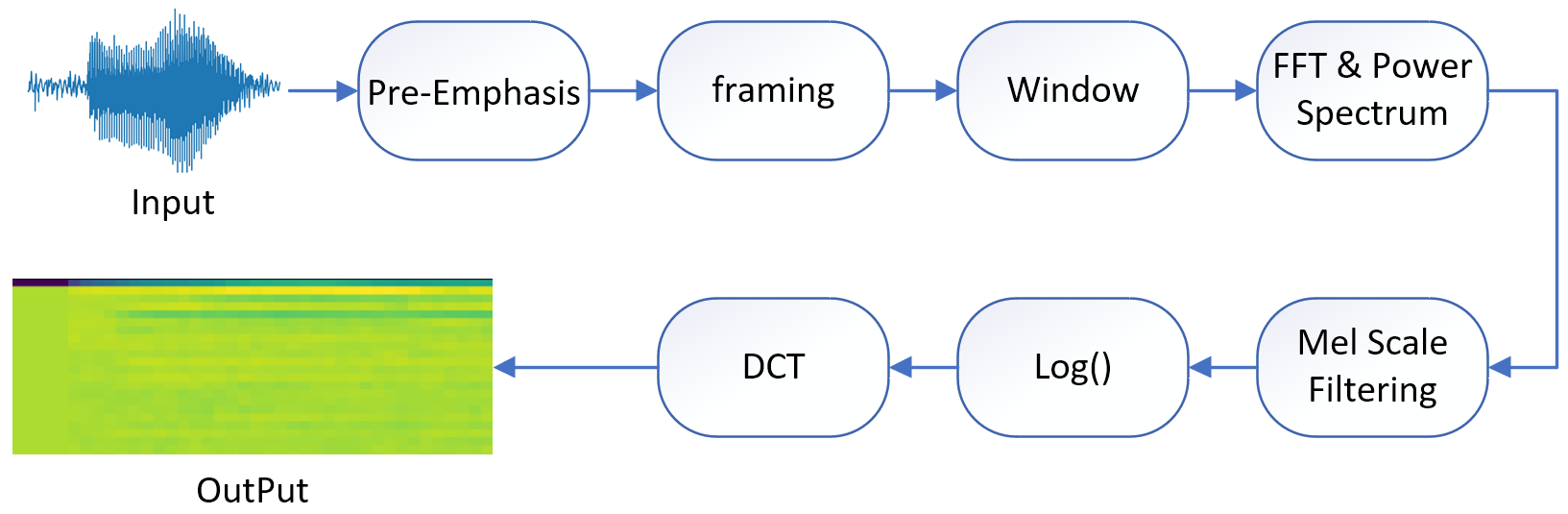}
\caption{MFCC block diagram~\cite{dave2013feature}.}\label{fig2}
\end{figure}

Figure~\ref{fig2} illustrates the MFCC process for feature extraction. The input audio undergoes \emph{pre-emphasis} to enhance high frequencies, followed by \emph{framing} and \emph{windowing} to reduce spectral leakage. The \emph{Short-Time Fourier Transform (STFT)} computes the power spectrum, which is processed through a \emph{triangular filter bank} to mimic human auditory perception. A \emph{Discrete Cosine Transform (DCT)} is then applied to the log-filtered energies to obtain MFCCs, retaining only relevant coefficients. Finally, \emph{mean normalization} ensures scale consistency across samples~\cite{dave2013feature,muda2010voice}.

For this study, MFCC is configured with a pre-emphasis parameter ($\alpha$) set to 0.97~\cite{dave2013feature}, a sample rate of 16,000 Hz, a frame size of 25 milliseconds, and 512 FFT points (NFFT), tailoring it to the specific characteristics of the audio data. To capture crucial acoustic information, both the MFCC and cepstral coefficients were set to 13, ensuring a comprehensive representation of the spectral content.

\subsection{The Proposed DNN Architecture}\label{sub2sec2}

\begin{figure*}[t]%
\centering
\includegraphics[width=0.8\textwidth]{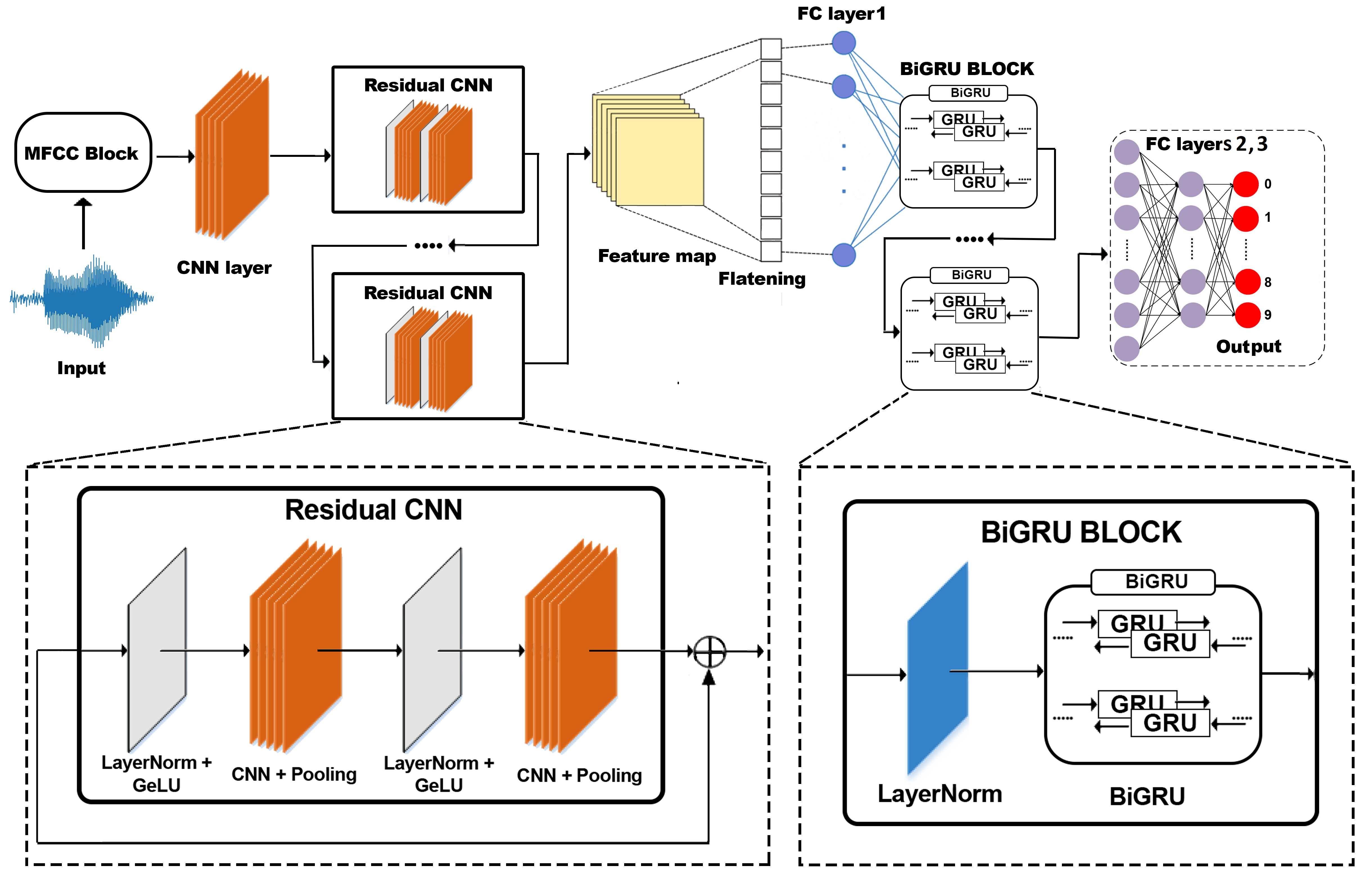}
\caption{Block diagram of the proposed DNN.}\label{fig3}
\end{figure*}

The proposed DNN for Persian digit recognition is inspired by DeepSpeech2~\cite{amodei2016deep}, with modifications to enhance performance. It consists of a CNN layer, residual CNN blocks, a Fully Connected (FC) layer, and Bidirectional GRU (BiGRU) blocks, as illustrated in Figure~\ref{fig3}. Each component plays a key role in feature extraction and classification.

The CNN layer~\cite{shahidi2022deep} extracts features and adjusts input dimensions. Next, three residual CNN blocks~\cite{he2016deep} refine feature representation. Unlike DeepSpeech2, word units are used instead of phonemes, and the Cross-Entropy Loss function replaces Connectionist Temporal Classification (CTC). The FC layer~\cite{basha2020impact} further adjusts feature dimensions before classification.
This layer, added specifically for the loss function, constitutes another difference from the DeepSpeech2 architecture. 

As opposed to the LSTM architecture used in DeepSpeech2, the proposed method employs BiGRU blocks~\cite{tao2019air}. This modification is motivated by GRU’s advantage of having fewer parameters than that of the LSTM, resulting in a reduction in the network weights and the computational resource consumption while often achieving superior performance. In addition, the BiGRU’s configuration enables the network to process each frame by considering data from both previous and next frames, leading to more accurate predictions. Five BiGRU blocks classify features, followed by two FC layers with a softmax function to determine results.

These modifications optimize DeepSpeech2 for robust Persian digit recognition in noisy environments. The following sections detail the CNN layer, residual CNN blocks, FC layer, and BiGRU blocks.

\subsubsection{CNN Layer}\label{subsub2sec2}

The CNN layer transforms input audio data into feature vectors while adjusting input dimensions. A single CNN layer is used with one input channel and 32 output channels to extract diverse features from the raw audio input. A kernel size of 3 is utilized to reinforce feature extraction, while a stride of 2 enables down-sampling, reducing computational complexity while preserving essential information.

\subsubsection{Residual CNN Blocks~\cite{he2016deep}}\label{sub2sub2sec2}

Three residual CNN blocks are used to learn audio features, enhancing the model’s depth and ability to recognize complex patterns in spoken Persian digits, even in noisy environments.

Each block includes two normalization layers (normalized shape: 20) to stabilize and accelerate the training process, two Gaussian Error Linear Unit (GELU) activation functions for contribute to the non-linearity of the model, and two CNN layers (32 input/output channels, kernel size: 3, stride: 1, padding: 1). Crucially, the output from these layers is summed with the input to form the final result of the Residual CNN block. This design choice enables the network to retain essential information from the input while incorporating the features extracted through the normalization and convolutional layers.

\subsubsection{FC Layer~\cite{basha2020impact}}\label{sub3sub2sec2}

An FC layer connects the residual CNN blocks to the BiGRU, optimizing extracted features before passing them forward. Configured with an input size of 25,600 and an output size of 512, it compresses information for efficient processing.

Additionally, two FC layers handle classification, each with dropout regularization, GELU, and softmax activation functions. The first FC layer reduces 1,024 inputs to 512 outputs, simplifying classification, while the second maps 512 inputs to 10 outputs. The final softmax activation converts outputs into probability scores across the 10 Persian digit classes, ensuring accurate recognition. This configuration allows for effective classification and prediction, contributing to the overall accuracy of our neural network in recognizing spoken Persian digits.

\subsubsection{BiGRU Block~\cite{tao2019air}}\label{sub4sub2sec2}

The BiGRU block, applied five times, captures temporal dependencies essential for recognizing spoken Persian digits. Each block includes layer normalization with a parameter of 512 to stabilize training and prevent vanishing gradient problem often associated with recurrent neural networks. A GELU activation function adds non-linearity to the model, enhancing its capacity to capture complex patterns.

At its core, the BiGRU processes input bidirectionally with an input size of 512 and a hidden layer size of 512. This design enables the network to process input sequences in both forward and backward directions, capturing contextual information from both past and future time steps.

\section{Results and Discussion}\label{sec4}

\begin{figure}[h]%
\centering
\includegraphics[width=0.5\textwidth]{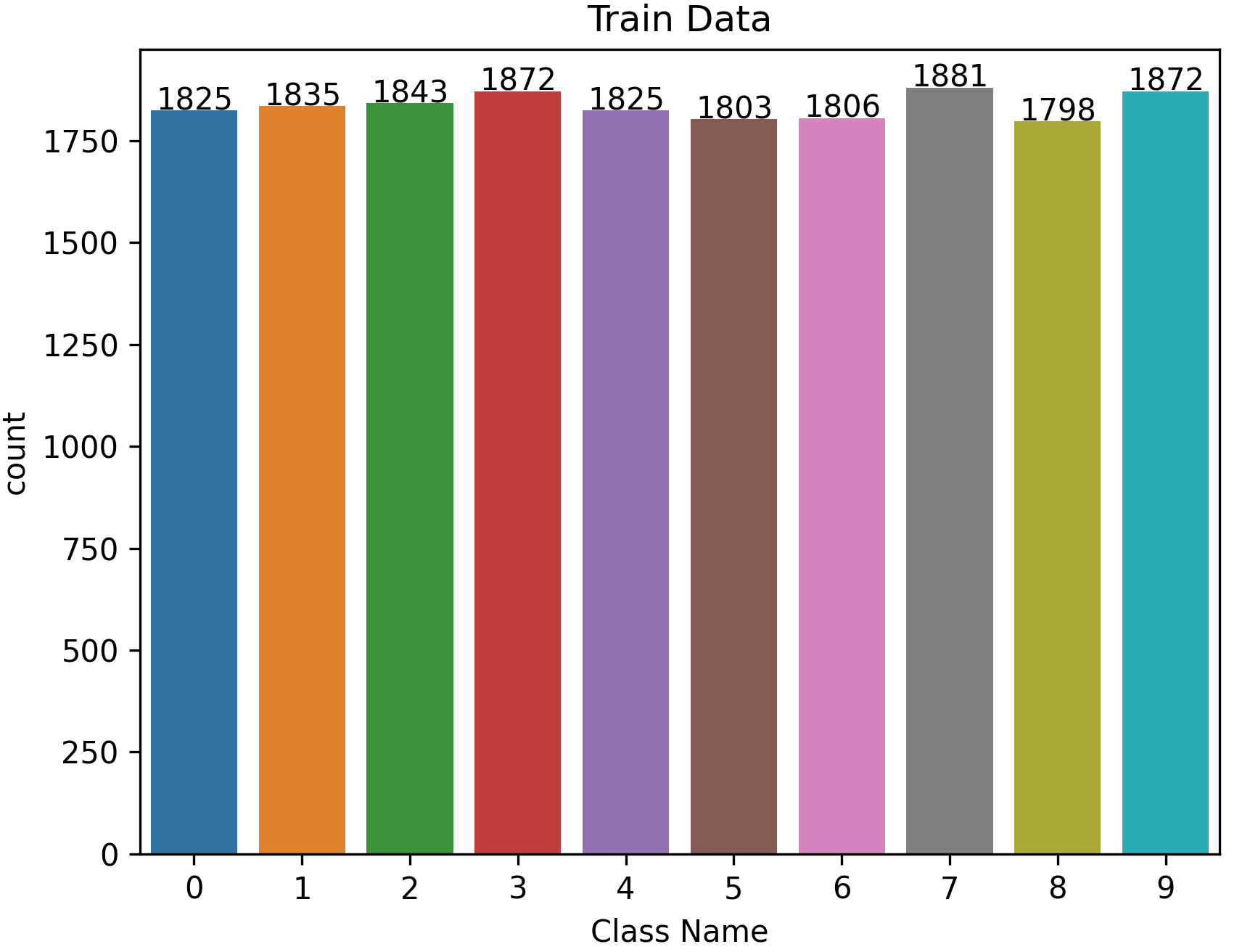}
\caption{Frequency chart of each class in Train data.}\label{fig4}
\end{figure}

\begin{figure}[h]%
\centering
\includegraphics[width=0.5\textwidth]{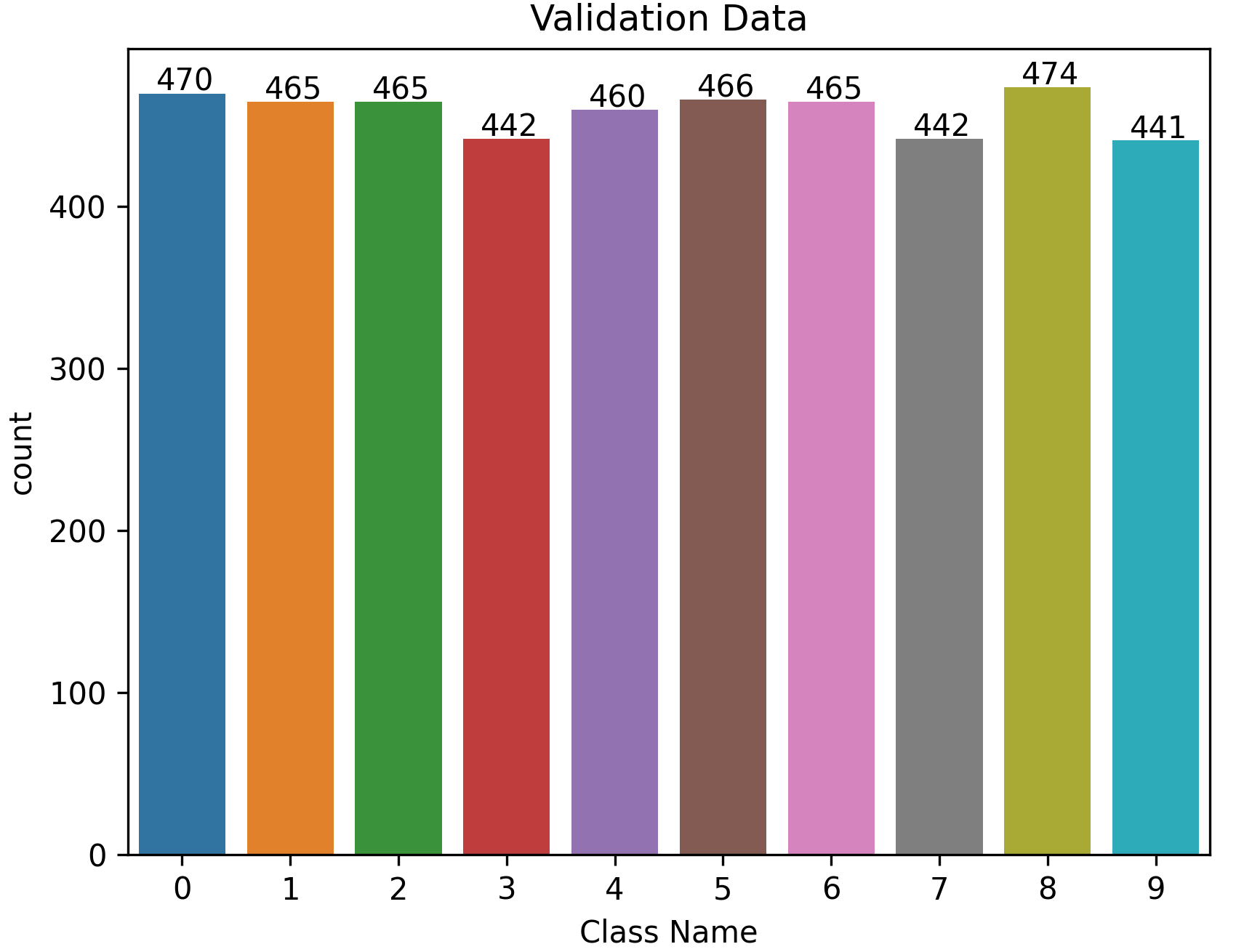}
\caption{Frequency chart of each class in the Validation data.}\label{fig5}
\end{figure}

\begin{figure}[h]%
\centering
\includegraphics[width=0.5\textwidth]{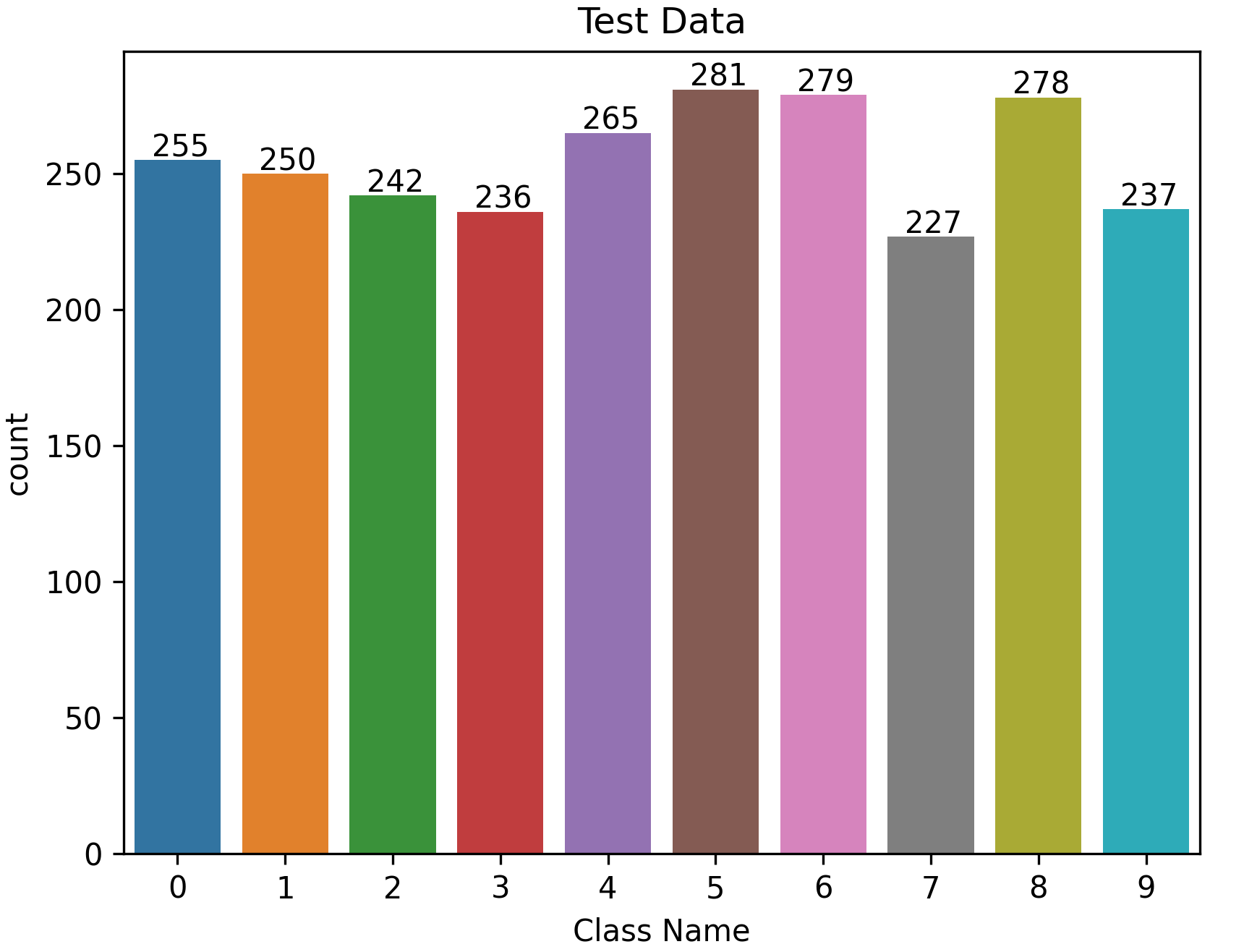}
\caption{Frequency chart of each class in the Test data.}\label{fig6}
\end{figure}

The dataset is divided into training, validation, and testing sets, with frequency distributions shown in Figures~\ref{fig4},~\ref{fig5}, and~\ref{fig6}, respectively. Data shuffling is performed before applying MFCC for feature extraction.

Training is conducted on Google Colab, utilizing a Tesla T4 GPU (15.1 GB RAM, 78.19 GB storage, 12.68 GB processing RAM). The model is implemented in Python programming language and the library used to create and train the model is PyTorch. 

Initially, simpler networks like CNN and GRU are evaluated but achieve limited validation accuracies of 83.22\% and 78.76\%, respectively, due to their smaller sizes and struggling to effectively learn from the noisy data. The LSTM model improves upon these, achieving 87.86\%.

\begin{figure}[h]%
\centering
\includegraphics[width=0.52\textwidth]{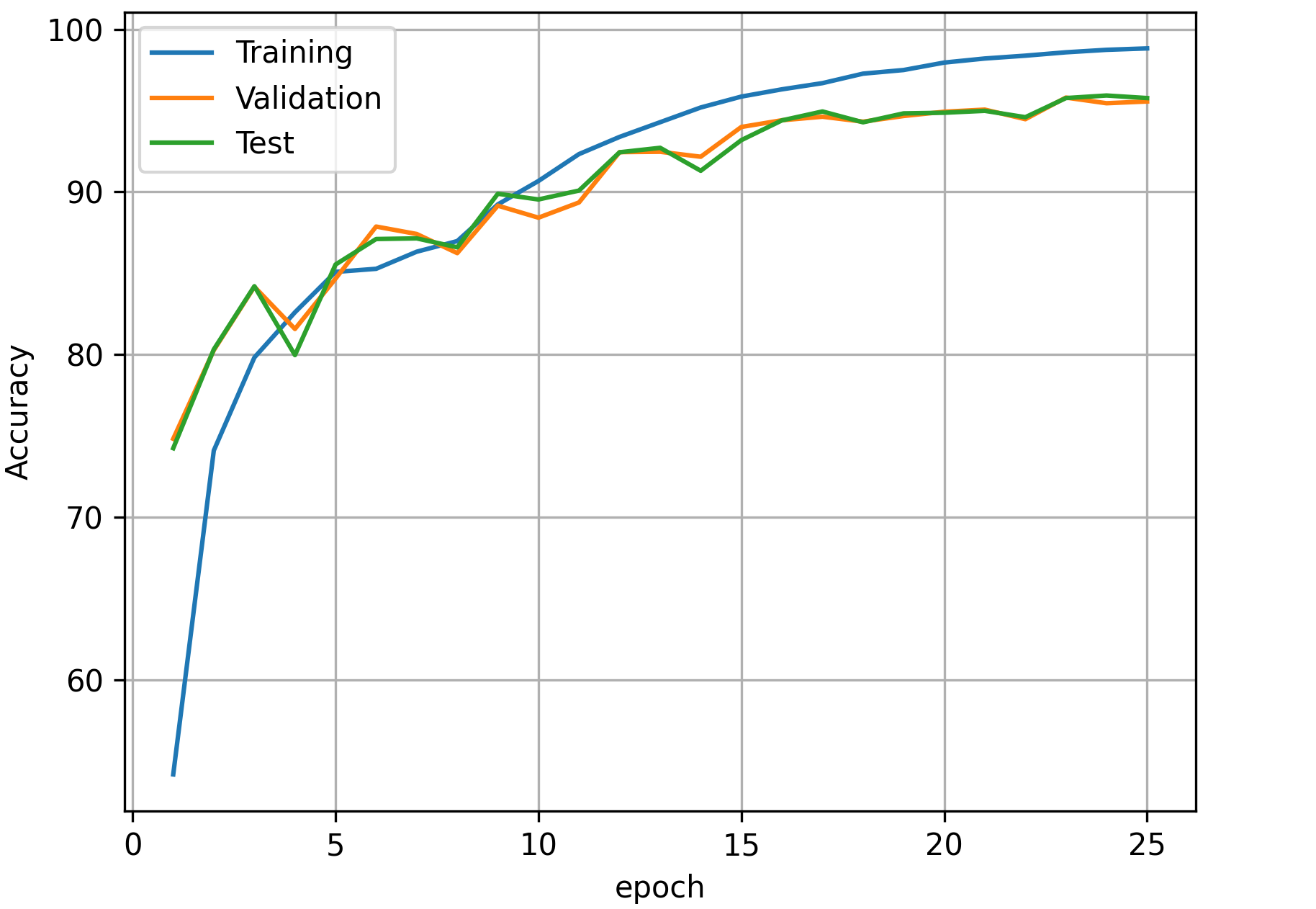}
\caption{Recognition accuracy chart on training, validation, and testing data.}\label{fig7}
\end{figure}

Figure~\ref{fig7} illustrates the proposed model’s accuracy across training, validation, and testing over 25 epochs. The final accuracy reaches 98.53\% (training), 96.10\% (validation), and 95.92\% (testing). Table~\ref{tab1} compares these results, confirming the superior performance of the proposed model.

\begin{table}[h]
\small 
\centering
\caption{Comparison of accuracy of LSTM, CNN, GRU, and the proposed DNN.}
\label{tab1}
\begin{tabular}{lccc}
\toprule
Network  & Training  & Validation  & Test  \\
\midrule
LSTM  & 91.16\%  & 87.45\%  & 86.82\%  \\
GRU   & 76.05\%  & 83.22\%  & 83.49\%  \\
CNN   & 82.90\%  & 80.34\%  & 80.59\%  \\
Proposed DNN & \textbf{98.53\%} & \textbf{96.10\%} & \textbf{95.92\%}  \\
\bottomrule
\end{tabular}
\end{table}

\begin{figure}[h]%
\centering
\includegraphics[width=0.5\textwidth]{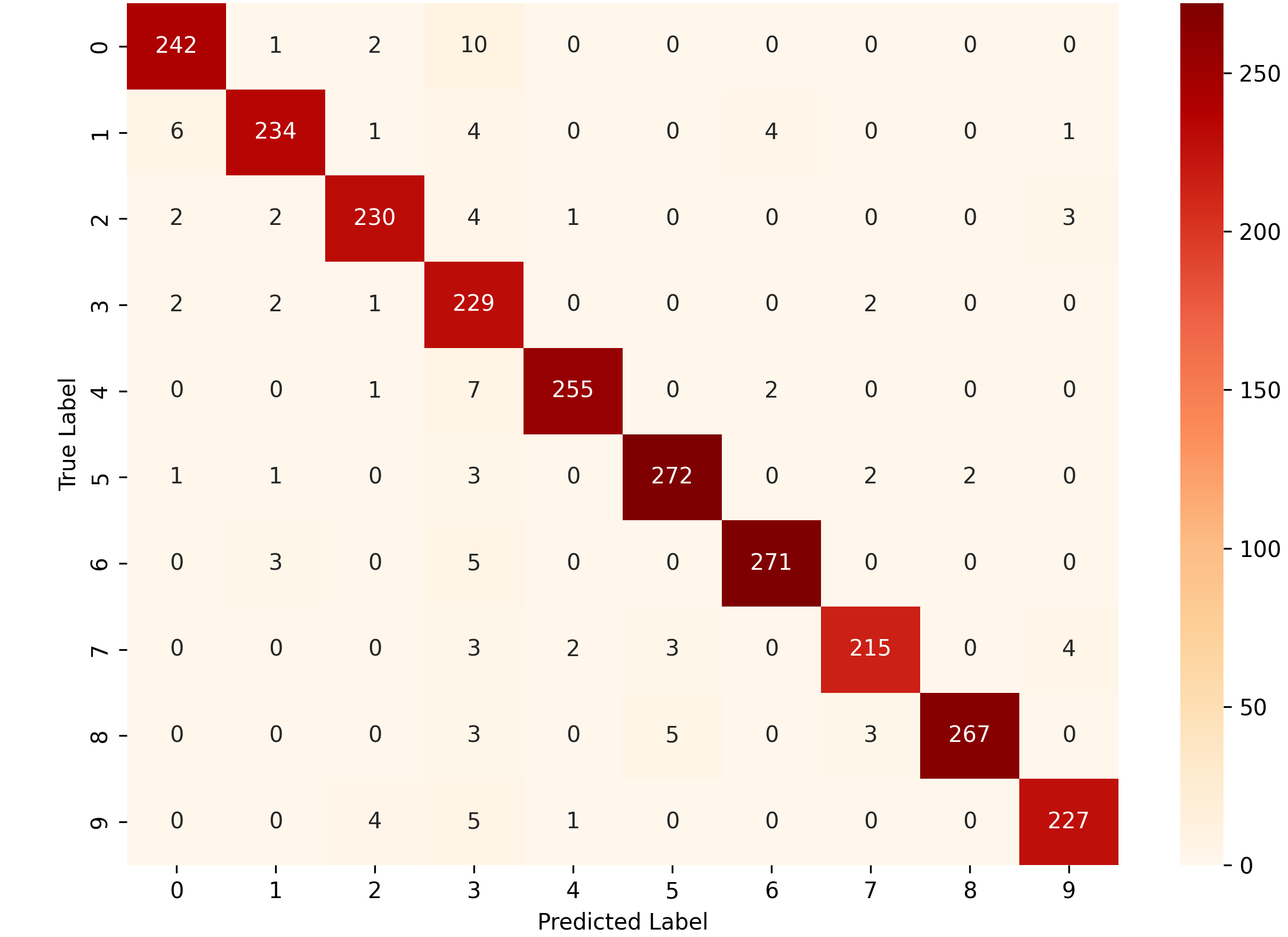}
\caption{Confusion matrix for the test set.}\label{fig8}
\end{figure}

Figure~\ref{fig8} presents the confusion matrix from the final test phase, indicating errors primarily in data with facet similarities. However, the error rate remains significantly lower than in phonologically based networks.

Table~\ref{tab2} compares the proposed method’s accuracy with related works.
Our approach outperforms the MTDRCC with MLP by 7.61\% and achieves a 26.88\% higher average accuracy than the phoneme unit-based LSTM method for Persian numbers in noisy conditions.

\begin{table}[h]
\centering 
\footnotesize 
\setlength{\tabcolsep}{5pt} 
\caption{Accuracy comparison of our method with related works.}\label{tab2}
\begin{tabular}{lll}
\toprule
\multirow{2}{5em}{Reference} & \multirow{2}{10em}{Method} & \multirow{2}{10em}{Accuracy (\%)} \\
\\
\arrayrulecolor{lightgray}\hline
\multirow{2}{5em}{2003~\cite{homayounpour2003rec}} & \multirow{2}{10em}{HMM + MLP (clean)}  & \multirow{2}{10em}{83.7 (cont. num.), 99.1 (disc. num.)} \\
\\
\hline
2016~\cite{dhanashri2017isolated} & DBN  & 86.06 (valid.)    \\
\hline
2020~\cite{zada2020pashto}  & MFCC + CNN &  84.17 (valid.) \\
\hline
\multirow{2}{5em}{2021~\cite{tabibian2021}}  &   \multirow{2}{10em}{LSTM (clean/noisy)}  & \multirow{2}{10em}{91.7 (clean), 69.2 (noisy)} \\
\\
\hline
\multirow{2}{5em}{2022~\cite{Z2N}}  &   \multirow{2}{10em}{MTDRCC + MLP}  & \multirow{2}{10em}{98.85 (clean), 88.49 (noisy)} \\
\\
\hline
\multirow{2}{5em}{2022~\cite{oruh2022long}}  & \multirow{2}{10em}{RNN + LSTM (clean)}  & \multirow{2}{10em}{99 (valid.)} \\
\\
\hline
\multirow{2}{5em}{2022~\cite{amadeus2022digit}} & \multirow{2}{10em}{Transfer Learning (AlexNet, GoogleNet)} & \multirow{2}{10em}{72 (AlexNet), 66 (GoogleNet)} \\
\\
\hline
\multirow{2}{5em}{Our Method} &  \multirow{2}{10em}{Residual CNN + BiGRU (noisy)} & \multirow{2}{10em}{98.53 (train), 96.10 (valid.)}\\
\\
\bottomrule
\end{tabular}
\end{table}

\section{Conclusion}\label{sec5}

This paper presents a deep neural network (DNN) for Persian spoken digit recognition, integrating CNN, residual CNN, BiGRU, and fully connected layers. By adopting word units instead of phoneme units, the model effectively handles facet similarities, enhancing feature extraction and recognition, particularly in noisy conditions. Experimental results confirm that the proposed DNN surpasses phoneme-based and LSTM-based methods in accuracy. This word unit-based approach offers a robust solution for spoken digit recognition, contributing to advancements in speech recognition technology with potential real-world applications.

\end{document}